\documentclass[twocolumn,twoside]{revtex4}
\usepackage{graphicx}
\usepackage{fancyhdr}
\def\beq{\begin{equation}}
\def\eeq{\end{equation}}
\def\bea{\begin{eqnarray}}
\def\eea{\end{eqnarray}}
\def\nn{\nonumber}
\def\nl{\nonumber\\}
\def\roughly#1{\mathrel{\raise.3ex\hbox
{$#1$\kern-.75em\lower1ex\hbox{$\sim$}}}}

\def\gsim{\roughly>}

\def\sla#1{\raise.15ex\hbox{$/$}\kern-.57em #1}% Feynman slash
\def\bsee{b \to s e^+ e^-}
\def\bsmumu{b \to s \mu^+ \mu^-}

\def\bsll{b \to s \ell^+ \ell^-}
\def\bctaunu{b \to c \,\tau^- {\bar\nu}_\tau}
\def\bcmunu{b \to c \,\mu^- {\bar\nu}_\mu}
\def\bclnu{b \to c \,\ell^- {\bar\nu}_\ell}
\renewcommand{\O}{\mathcal{O}}
\newcommand{\hc}{\mathrm{h.c.}}
\def \cB{{\cal B}}

\begin{document}

\preprint{UdeM-GPP-TH-22-xxx}

%Title of paper
\title{\boldmath The $B$ Anomalies and non-SMEFT New Physics}

% Repeat the \author .. \affiliation  etc. as needed
%
% \affiliation command applies to all authors since the last
% \affiliation command. The \affiliation command should follow the
% other information

\author{David London}
\affiliation{Physique des Particules, Universit\'e de Montr\'eal, \\
1375 Avenue Th\'er\`ese-Lavoie-Roux, Montr\'eal, QC, Canada  H2V 0B3}

\begin{abstract}
The modern viewpoint is that the Standard Model is the leading part of an effective field theory that obeys the symmetry $SU(3)_C \times SU(2)_L \times U(1)_Y$. Since the discovery of the Higgs boson, it is generally assumed that this symmetry is realized linearly (SMEFT), but a nonlinear realization (e.g., HEFT) is still possible. The two differ in their predictions for the size of certain low-energy dimension-6 four-fermion operators: for these, HEFT allows $O(1)$ couplings, while in SMEFT they are suppressed by a factor $v^2/\Lambda_{\rm NP}^2$, where $v$ is the Higgs vev. In this talk, I argue that (i) such non-SMEFT operators contribute to both $\bsll$ and $\bctaunu$, transitions involved in the present-day $B$ anomalies, (ii) the contributions to $\bsll$ are constrained to be small, at the SMEFT level, and (iii) the contribution to $\bctaunu$ can be sizeable. I show that the angular distribution in ${\bar B} \to D^* (\to D \pi') \, \tau^{-} (\to \pi^- \nu_\tau) {\bar\nu}_\tau$ contains enough information to extract the coefficients of all new-physics operators. The measurement of this angular distribution can tell us if non-SMEFT new physics is present.
\end{abstract}

%\maketitle must follow title, authors, abstract
\maketitle

\thispagestyle{fancy}

% body of paper here - Use proper section commands
% References should be done using the \cite, \ref, and \label commands
% Put \label in argument of \section for cross-referencing
%\section{\label{}}

\section{SMEFT vs.\ HEFT}

The Standard Model (SM) of particle physics explains almost
all experimental data to date. There is no doubt that it is correct. This said, the SM is not complete: for example, it has no explanation for neutrino masses, dark
matter and the baryon asymmetry of the universe. There must
exist physics beyond the SM. And since no new particles have been seen
at the LHC, this new physics (NP) must be heavy.

The modern view is that the SM is just the leading part of an
effective field theory (EFT) (see, 
  e.g., Refs.~\cite{EFTBook, Burgess:2007pt}), obtained when this NP is integrated out
at a scale of $O({\rm TeV})$). This EFT must respect the SM symmetry
group $SU(3)_C \times SU(2)_L \times U(1)_Y$.

But this raises the question: is the SM symmetry realized linearly
[SMEFT (see, e.g.,
Refs.~\cite{Buchmuller:1985jz, Brivio:2017vri})] or nonlinearly [e.g., HEFT (see, 
  e.g., Refs.~\cite{Feruglio:1992wf, Bagger:1995mk, Burgess:1999ha,
  Grinstein:2007iv, Buchalla:2012qq, Alonso:2012px, Alonso:2012pz,
  Buchalla:2013rka, Cohen:2020xca})]? Since the discovery of the Higgs
boson, the SMEFT is the default assumption, but HEFT is still
possible. 

The question of whether the symmetry is realized linearly
or nonlinearly
can only be answered experimentally. To see how this can be done, one must compare power counting in SMEFT and in HEFT.

Consider a non-standard $Z{\bar u}_R u_R$
coupling, $g_z \, Z_\mu (\bar u \gamma^\mu P_R u)$. Within HEFT, it
has mass-dimension 4, so $g_z \sim O(1)$. But within SMEFT, it arises
at dimension 6 from $\Lambda_h^{-2} (H^\dagger D_\mu H) (\bar u
\gamma^\mu P_R u)$. Thus, $g_z \sim v^2/\Lambda_h^2$.
There are other dimension-6 operators that do not involve the Higgs
field (e.g., 4-fermion operators). Their coefficients $\sim
\Lambda^{-2}$, where $\Lambda$ is the scale of NP, $\gsim O({\rm
  TeV})$.
Within SMEFT, the assumption is that $\Lambda_h = \Lambda$. This
implies that SMEFT predicts that $g_z$ is considerably smaller than
the value allowed by HEFT.

In order to test the SMEFT assumption, we must (i) identify operators whose
power counting is different in SMEFT and HEFT, and (ii) find ways of
measuring these operators. If it is found that the coefficient of such
an operator is larger than what is predicted by SMEFT, this points to
non-SMEFT NP.

\section{LEFT}

As the title of the talk indicates, we want to test SMEFT with $B$ decays. At this energy scale, $O(m_b)$, the EFT is the LEFT (low-energy effective field theory, also known as WET, weak
effective field theory), obtained when the heavier SM particles ($W^\pm$, $Z^0$, $H$, $t$) are also
  integrated out. (This is like the Fermi theory.)

In Ref.~\cite{Jenkins:2017jig}, a complete and non-redundant basis of LEFT operators up to dimension 6 is presented. We focus on those operators that conserve lepton and baryon number. All dimension-6 LEFT operators must respect $U(1)_{em}$. Most of them are also invariant under $SU(2)_L \times U(1)_Y$, and can be generated from dimension-6 SMEFT operators. However, a handful of dimension-6 LEFT operators are not invariant under $SU(2)_L \times U(1)_Y$. That is, they are not generated by dimension-6 SMEFT operators. These ``non-SMEFT operators'' are the ones that interest us.

In Ref.~\cite{Burgess:2021ylu}, my collaborators and I examine the LEFT operators and identify eleven types of non-SMEFT four-fermion operators: there are six four-quark operators, one four-lepton operator, and four (semileptonic) operators with two quarks and two leptons. 

But this is just the first step; we must also find ways of measuring them.

\newpage

\section{\boldmath The $B$ Anomalies}

At the present time, there are discrepancies with the SM in measurements of a number of observables involving the semileptonic transitions $\bsll$ ($\ell = e,~\mu)$ and $\bctaunu$ decays. These are the $B$ anomalies (for a review, see Ref.~\cite{London:2021lfn}). Our list of non-SMEFT four-fermion operators includes contributions to these two types of decays. It is only natural to explore whether these can be measured in tests of the $B$ anomalies. 

There are two dimension-6 non-SMEFT operators that contribute to $\bsll$. Along with the dimension-8 SMEFT operators to which they are mapped at tree level, they are
\bea 
&& \O_{ed}^{S,RR} \equiv (\overline e_{Lp}e_{Rr})(\overline d_{Ls}d_{Rt}) \nl 
&& ~~~~ \longrightarrow
Q_{\ell eqdH^2}^{(3)} \equiv (\overline \ell_pe_rH)(\overline q_sd_tH) ~, \nl
&& \O_{ed}^{T,RR} \equiv (\overline e_{Lp}\sigma^{\mu\nu}e_{Rr})(\overline d_{Ls}\sigma_{\mu\nu}d_{Rt}) \nl
&& ~~~~ \longrightarrow
Q_{\ell eqdH^2}^{(4)} \equiv (\overline \ell_p\sigma_{\mu\nu}e_rH)(\overline q_s\sigma^{\mu\nu}d_tH) ~.
\eea
Here, the $\O$s are dimension-6 four-fermion LEFT operators \cite{Jenkins:2017jig}, while the $Q$s are dimension-8 SMEFT operators, taken from Ref.~\cite{Murphy:2020rsh}. 
The subscripts $p, r, s, t$ are generation indices; they are suppressed in the operator labels. In the LEFT operators, the chiralities of $e$ and $d$ are given explicitly, while in the SMEFT operators, $\ell$ and $q$ denote left-handed $SU(2)_L$ doublets, and $e$, $u$ and $d$ denote right-handed $SU(2)_L$ singlets. 

Both of these LEFT operators contribute to $\bsmumu$. However, they contribute in a chirally unsupressed way, unlike the SM. Because of this, the addition of either of these operators can dramatically change the prediction for $\cB(B_s^0 \to \mu^+ \mu^-)$. But the present measured value is $\cB(B_s^0 \to \mu^+ \mu^-) = (2.9 \pm 0.4) \times 10^{-9}$ \cite{ParticleDataGroup:2020ssz},
close to the SM prediction. We find that the coefficients of these non-SMEFT
operators cannot be larger than $O(10^{-4})$, which is consistent with
SMEFT expectations. Things are similar for the analogous operators that
contribute to $\bsee$: the upper limit of $\cB(B_s^0 \to
e^+ e^-) < 9.4 \times 10^{-9}$ \cite{ParticleDataGroup:2020ssz}
constrains the coefficients of these operators to be $< O(10^{-3})$,
again consistent with SMEFT. We therefore conclude that there is no indication that non-SMEFT NP contributes to $\bsll$.

The non-SMEFT (dimension-6) four-fermion operator that contributes to $\bclnu$ is 
\beq
\O_{\nu edu}^{V,LR} \equiv (\overline\nu_{Lp}\gamma^\mu e_{Lr})(\overline d_{Rs}\gamma_\mu u_{Rt}) + \hc 
\label{OVLR}
\eeq
It maps to SMEFT operators as follows:
\bea
&\longrightarrow& 
Q_{Hud} \equiv i(\tilde H^\dagger D_\mu H)(\overline u_p\gamma^\mu d_r) + \hc \nl
&& ~~~~~ ({\hbox{dimension 6}}) ~, \nl
&\longrightarrow& 
Q_{\ell^2udH^2} \equiv (\overline \ell_pd_rH)(\tilde H^\dagger \overline u_s\ell_t) + \hc \nl
&& ~~~~~ ({\hbox{dimension 8}}) ~, 
\eea
where $\tilde H = i \sigma_2 H^*$. Now, the fact that there is a mapping to the dimension-6 SMEFT operator $Q_{Hud}$ seems to suggest that $\O_{\nu edu}^{V,LR}$ is itself a SMEFT operator after all. But one must be careful here: $Q_{Hud}$ is a lepton-flavour-universal operator that generates equal effective couplings for the operators $\O_{\nu \ell b c}^{V,LR}$, $\ell = e$, $\mu$, $\tau$ \cite{Bernard:2006gy}. But we are interested in an 
effective operator that generates {\it only} $\O_{\nu \tau b
  c}^{V,LR}$ without the other two, i.e., it violates lepton-flavour
universality. This can be generated only by the dimension-8 SMEFT operator $Q_{\ell^2udH^2}$ given above.

Consider the decay $\bctaunu$.  Assuming only left-handed neutrinos,
the most general LEFT effective Hamiltonian describing this decay involves
five four-fermion operators:
\bea
    {\cal H}_{eff} &=& \frac{4 G_F}{\sqrt{2}} \, V_{cb} \, O_V^{LL} 
- \frac{C_V^{LL}}{\Lambda^2} \, O_V^{LL} \, - \frac{C_V^{LR} }{\Lambda^2} \,
 O_V^{LR} \,, \nn \\
 && -~\frac{C_S^{LL} }{\Lambda^2} \, O_S^{LL} \,- \frac{C_S^{LR} }{\Lambda^2} \, O_S^{LR}
 - \frac{C_T }{\Lambda^2} \, O_T ~,
\eea
with
\bea
O_V^{LL,LR} &\equiv& ({\bar \tau} \gamma^\mu P_L \nu ) \, ( {\bar c} \gamma_\mu P_{L,R} b ) ~, \nl
O_S^{LL,LR} &\equiv& ({\bar \tau} P_L \nu )\, ({\bar c} P_{L,R} b) ~, \nn\\
O_T &\equiv& ({\bar \tau} \sigma^{\mu\nu}  P_L \nu )\, ({\bar c} \sigma_{\mu\nu} P_L b) ~. 
\eea

\null
\vskip-5truemm
\noindent
But note: $O_V^{LR}$ is the non-SMEFT operator of Eq.~(\ref{OVLR}); its coefficient $C_V^{LR}$ is suppressed by $v^2/\Lambda^2$ in SMEFT. For this reason, it is usually ignored when looking for NP in $\bctaunu$.

For example, fits to the $\bctaunu$ data are performed in Refs.~\cite{Blanke:2018yud, Blanke:2019qrx}, but with $O_V^{LR}$ excluded (precisely because it is a non-SMEFT operator). In Ref.~\cite{Burgess:2021ylu}, we redo the fit, including $O_V^{LR}$. The results are shown in Table \ref{fitresults}. Although the best fit is provided by the scenario in which  only $C_V^{LL}$ is added, the fit remains acceptable when both $C_V^{LL}$ and $C_V^{LR}$ are allowed to be nonzero. And in this scenario, the best fit has $C_V^{LR} = O(1)$. Such a large value is allowed within HEFT, but not SMEFT. 

\begin{table}[h]
%\begin{adjustbox}{width=0.45\textwidth}
\caption{Fit results for the scenarios in which $C_V^{LL}$, $C_V^{LR}$
  or both $C_V^{LL}$ and $C_V^{LR}$ are allowed to be nonzero.}
\begin{tabular}{|c|c|c|}
  \hline
  New-physics coeff.\ & Best fit & $p$ value (\%) \\
\hline
\hline
$C_V^{LL}$ & $-3.1 \pm 0.7$ & 51 \\
$C_V^{LR}$ & $2.8 \pm 1.2$ & 0.3 \\
$(C_V^{LL},C_V^{LR})$ & $(-3.0 \pm 0.8,0.6 \pm 1.2)$ & 35 \\
\hline
\end{tabular}
%\end{adjustbox}
\label{fitresults}
\end{table}

The bottom line is that a non-SMEFT value of $C_V^{LR}$ is allowed by the present data. Furthermore -- and this is the point of this talk -- it can be measured.

\section{Measuring Non-SMEFT Operators}

Consider ${\bar B} \to D^* \tau^{-} {\bar\nu}_\tau$. Because ${\bar B}$ is a pseudoscalar meson, while $D^*$ is a vector, there are only four NP operators that contribute: $O_V^{LL}$, $O_V^{LR}$, $O_{LP} \equiv O_S^{LR} - O_S^{LL}$, $O_T$. This implies that there are seven NP parameters  (four magnitudes, three relative phases).

In Ref.~\cite{Bhattacharya:2020lfm}, it is proposed to measure the angular distribution in ${\bar B} \to D^* (\to D \pi') \, \tau^{-} (\to \pi^- \nu_\tau) {\bar\nu}_\tau$. The differential decay rate is a function of five parameters: $q^2$ (the momentum$^2$ of the $\tau^{-} {\bar\nu}_\tau$ pair), $E_\pi$ (the energy of the pion in $\tau^{-} \to \pi^- \nu_\tau$), and the three angles $\theta^*$, $\theta_\pi$, $\chi_\pi$ (see Fig.~I). The data is separated into $q^2$-$E_\pi$ bins, and an angular analysis is performed in each bin.

\begin{figure}
\label{fig:angles}
\begin{center}
\includegraphics[width=0.47\textwidth]{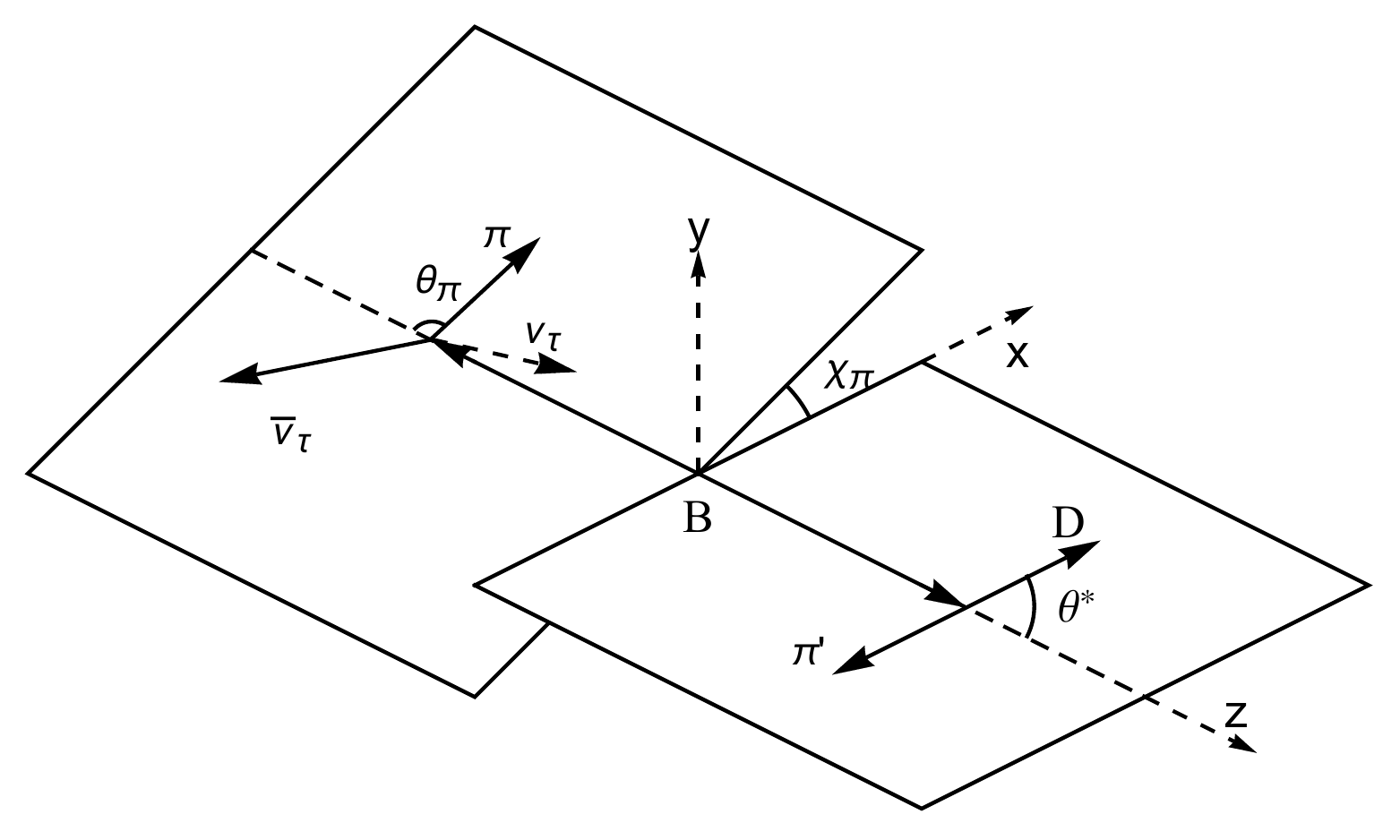}
\caption{Definition of the angles in the \\ $\bar{B} \to D^* (\to D \pi)
  \, \tau^- (\to \pi^- \nu_\tau) \bar{\nu}_{\ell}$ distribution.}
\end{center}
\end{figure}

The angular distribution can be written as
\bea
&& \sum^{9}_{i=1} f^R_i(q^2,E_\pi)\Omega^R_i(\theta^*,
\theta_\pi,\chi_\pi) \nl 
&& ~~~~~~~ + \sum^{3}_{i = 1}f^I_i(q^2,E_\pi)\Omega^I_i(\theta^*,
\theta_\pi,\chi_\pi) ~.
\eea
The nine $f^R_i \Omega^R_i$ terms are CP-conserving and are present in
the SM. The 3 $f^I_i \Omega^I_i$ terms are CP-violating.

The key point is the following. There are twelve observables (angular functions) in each $q^2$-$E_\pi$ bin. (And there may well be several bins -- the exact number will be decided by experiment.) But there are only seven NP parameters. This implies that all NP parameters can be extracted. If the value of $|C_V^{LR}|$ is found to be larger than that predicted by SMEFT, then not only will NP have been discovered, but it will have been determined that this is new, non-SMEFT NP.

Finally, we note that Ref.~\cite{Bobeth:2021lya} argues that there are hints of NP in $\bcmunu$. As with $\bctaunu$, there are several dimension-6 four-fermion NP operators that contribute to $\bcmunu$, including one non-SMEFT operator. This decay can be analyzed similarly to $\bctaunu$: the angular distribution for $\bcmunu$ described in Ref.~\cite{Bhattacharya:2019olg} provides enough observables to fit for the coefficients of all dimension-6 NP operators, including the non-SMEFT one.

\section{Recap}

It is generally thought that the SM is the leading part of an EFT produced when the heavy NP is integrated out. Since the discovery of the Higgs boson, this EFT is usually taken to be the SMEFT. However, this is an assumption; the identity of the EFT must be determined experimentally.

One difference between HEFT and SMEFT is power counting: the coefficients of certain low-energy four-fermion operators are predicted to be considerably smaller in SMEFT than in HEFT. This suggests a simple test: find such an operator and measure its coefficient. If the value of the coefficient is found to be larger than that predicted by SMEFT, this imples that the NP operator is of non-SMEFT type. 

At present, there are anomalies in $\bsll$ and $\bctaunu$ decays. Both of these receive contributions from non-SMEFT NP operators. In the case of $\bsll$, the coefficients of such operators are constrained to be small, of SMEFT size. But the coefficient of the non-SMEFT $\bctaunu$ operator $O_V^{LR}$ is still allowed by present data to be sizeable. 

The angular distribution in ${\bar B} \to D^* (\to D \pi') \, \tau^{-} (\to \pi^- \nu_\tau) {\bar\nu}_\tau$ provides enough observables to extract all NP parameters from a fit. If the coefficient of $O_V^{LR}$ is found to be larger than the SMEFT prediction, this will point to the presence of non-SMEFT NP.

% If you have acknowledgments, this puts in the proper section head.
%\bigskip % extra skip inserted
\begin{acknowledgments}
I thank Cliff Burgess, Serge Hamoudou and Jacky Kumar for collaboration on Ref.~\cite{Burgess:2021ylu}.
This work was partially financially
supported by funds from the Natural Sciences and Engineering Research
Council (NSERC) of Canada. 
\end{acknowledgments}

\bigskip % extra skip inserted
% Create the reference section using BibTeX:
%\bibliography{basename of .bib file}
%\begin{thebibliography}{9}   % Use for  1-9  references

\end{document}